\documentclass[noshowpacs,twocolumn,preprintnumbers,secnumarabic,amsmath,amssymb,nofootinbib]{revtex4}
\usepackage{dcolumn}     
\usepackage{bm}           
\usepackage{graphicx}

\usepackage{amsmath,amssymb}  
\usepackage{bm}  
\usepackage[pdftex,bookmarks,colorlinks,breaklinks]{hyperref}  
\usepackage{mathrsfs}
\usepackage{amsmath}
\usepackage{bm}

\begin{document}
\title{The decoupling of scalar-modes from a linearly perturbed dust-filled Bianchi type I model}
\author{Bob  Osano}
\address{ Cosmology and Gravity Group, Department of Mathematics and Applied Mathematics,
University of Cape Town, Rondebosch, 7701, South Africa}
\date{\today}
\email{bob.osano@uct.ac.za} 
\date{August 15, 2013 }

\begin{abstract}
We study linear perturbations about a dust filled Bianchi Type I model with the vorticity set to zero. In comparison to linear perturbations about FLRW models, modes of perturbations about Bianchi type I models are coupled. We find that the tensor that represents the background shear needs to be degenerate in order  for the scalar-mode perturbations to decouple from the rest of the flow.
\end{abstract}

\maketitle
 PACS: 04.20.Ex, 04.20.-q, 98.80.jk, 98.80.-k
\\
\par What we currently know about our physical universe is based on the analysis of a model that is expanding, isotropic and homogeneous, and that has a cosmological constant. This, together with the analysis of itÕs linear perturbations, undergird our picture of the universe \cite{bib:1}. The model accounts for the late time universe as is evident in the analysis of large scale cosmic microwave background observations \cite{bib:1,bib:2}. Although parameter determination from the analysis of CMB fluctuations appears to confirm this picture \cite{bib:10}, further analyses indicates that there are still anomalies \cite{bib:1}-\cite{bib:11}. In particular, it would appear that the universe may have a preferred direction \cite{bib:10.1}. It is not yet clear if the directional preference is intrinsic to the underlying model, and if so what implications this would has on modern cosmology. Recently released Planck results \cite{bib:Planck} indicate that these anomalies persists.
\par It was shown that some of the anomalies could be eliminated when WMAP first-year data was corrected with Bianchi $IV_{h}$ template \cite{bib:11}. The authors also find that the cosmological parameters required to reproduce the Bianchi morphology are inconsistent with what they called the "cosmic concordance" of other data sets and concluded that Bianchi models could not be a realistic physical explanation for these anomalies.  What is curious about these findings is that Bianchi Models appear ruled out because they are incompatible with {\it isotropic} inflationary models and by the 'cosmic no-hair conjecture' \cite{bib:12}. 
We need to point out that the proof of this conjecture, a conjecture that forms the basis for the above analysis, assumes the weak energy condition. With the advent of inflationary models, the violation of the energy conditions opens up other possibilities. In particular, it is worth investigating anisotropic inflationary models \cite{bib:13} as they may address the incompatibility between Bianchi models and standard inflationary models. Given that the proof of cosmic conjecture subject to weak energy condition and the anomaly in the WMAP data, it may be too soon to discard Bianchi models. This is the motivation for this letter. We seek to answer a slightly different question and one that is related to the Cauchy problem in General Relativity. The problem has to do with the preservation of the constraints equations subject to the propagation equations, recalling that the solutions to the constraints equations provide us with the initial conditions for the set of evolution equations. 

We use the covariant approach which differs from the standard metric perturbations approach in the sense that one does not perturb the metric and but a set of covariant nonlinear equations which are then linearized about a background of choice. The covariant equations are the result of splitting Einstein-Ricci-Bianchi equations, covariantly, into propagation and constraint equations \cite{bib:222}. The method has been applied to studies of different kinds of perturbations in an almost-FLRW model \cite{bib:21}. In this letter we want to apply it to density perturbations about Bianchi Type I model with dust equation of state, where the vorticity vanishes. Previous studies of perturbations of anisotropic models using alternative approaches include \cite{bib:23, bib:24, bib:20, bib:26, bib:35}. We have to address the following issue: immediately confronted with a major problem: 
\begin{itemize}
\item How does one characterize decoupled density perturbations about Bianchi I model given that the scalar, vector and tensor modes are coupled for perturbations about this background ? 
\item How do we know that the conditions governing our characterization are preserved?
\end{itemize} 
In order to answer these questions we need to define the scope of the problem at hand and to identify parameters that play a role in the problem. 

For a general spacetime given by a well defined manifold and its accompanying metric ($\mathcal{M},\textbf{g}$), one can assume that there is a well-defined preferred motion of matter which gives rise to a unique 4-velocity. This velocity may be given by the average motion of matter for a family of preferred. Mathematically, the 4-velocity, which we denoted by $u^{a}$ (\( =\frac{dx^{a}}{d\tau},\)) where $\tau$ is the proper time measured along the fundamental world-lines. The 4-velocity has the property \(u_{a}u^{a}=-1\) \cite{bib:22}. Physically, this velocity may be defined by the vanishing of the dipole of the cosmic microwave background radiation (CMB). Only one 4-velocity will set this dipole to zero \cite{bib:22}. Using the 4-velocity, we can then define, and make use of the following projection tensors. Parallel projector:
\begin{eqnarray}{U^{a}}_{b} &=&-u^{a}u_{b},\end{eqnarray} and with the properties
\begin{eqnarray}
{U^{a}}_{c}{U^{c}}_{b}={U^{a}}_{b},
{U^{a}}_{a}=1,~~~ U_{ab}u^{b}=u_{a},\end{eqnarray} 
Orthogonal projector: 
\begin{eqnarray}
h_{ab}&=&g_{ab}+u_{a}u_{b},
\end{eqnarray} which has the properties; \begin{eqnarray}{h^{a}}_{c}{h^{c}}_{b}={h^{a}}_{b}, ~~~{h^{a}}_{a}=3,~~~h_{ab}u^{b}=0.
\end{eqnarray}  We also make use of the permutation tensor $\varepsilon_{abc}$, which is given by 
\begin{eqnarray}
\varepsilon_{abc}=u^{d}\eta_{dabc}\Rightarrow\eta_{abc}=\eta_{[abc]},
\eta_{abc}u^{c}=0,
\end{eqnarray} and where $\eta_{abcd}$ is a 4 - dimensional volume element in the rest-space of the fundamental observer. We denote the covariant time derivative of the tensor ${T^{ab}}_{cd}$ by
\begin{eqnarray}
\dot{T^{ab}}_{cd}=u^{e}\nabla_{e}{T^{ab}}_{cd}.
\end{eqnarray} This represents derivative along the observer word-line. The orthogonally projected covariant derivative will be denoted by $\tilde{\nabla}_{a}$, so that the orthogonal projection a ${T^{ab}}_{cd}$ takes the form \[\tilde{\nabla}_{e}{T^{ab}}_{cd}={h^{a}}_{f}{h^{b}}_{g}{h^{p}}_{c}{h^{q}}_{d}{h^{r}}_{e}{\nabla}_{r}{T^{fg}}_{pq},\] where there is total projection on all indices. If the vorticity is set to zero, $\tilde{\nabla}$ coincides with the 3-dimensional covariant derivatives. Using these projection tensors, covariant derivatives, and symmetric properties, the 4-velocity $u_{a}$ may be split into the following irreducible parts:\begin{eqnarray}
\nabla_{a}u_{b} &=&
-u_{a}\dot{u}_{b}+\tilde{\nabla}_{a}u_{b}=\small\frac{1}{3}\Theta
h_{ab}+\sigma_{ab}+ \omega_{ab},
\end{eqnarray} where the trace $\Theta=\tilde{\nabla}_{a}u^{a}$ is
the volume-rate of expansion. This rate also determines the Hubble parameter; $H=\small{\frac{1}{3}}\Theta.$ $\sigma_{ab}=\tilde{\nabla}_{\langle a}u_{b\rangle}$ is the trace-free symmetric rate of shear tensor such that
$\sigma_{ab}u^{b}=0, \sigma_{a}^{a}=0.$ $\omega_{ab}=\tilde{\nabla}_{[a}u_{b]}$ is the skew-symmetric vorticity tensor. 
\par Orthogonal projection allows the above covariant quantities to represent respective behaviour in the rest-frame of the fundamental observer. This rest frame coincides with the orthogonal hyper-surfaces in the case where vorticity is absent. In general the Einstein-Ricci-Bianchi equations may be covariantly split into propagation and constraint equations. The full non-linear set of equations include, the propagation equations:
\begin{eqnarray}
\label{p:1}\dot{\rho}+\Theta\rho&=&0,\\
\label{p:2}\dot{\Theta}+\frac{1}{3}{\Theta^{2}}+\sigma_{ab}\sigma^{ab}+\frac{1}{2}\rho&=&0,\\
\label{p:3}\dot{\sigma}_{ab}+\frac{2}{3}\Theta\sigma_{ab}+{\sigma_{c\langle a}}{\sigma_{b\rangle}}^{c}+E_{ab}&=&0,\\
\label{p:4}\dot{E}_{ab}+\Theta E_{ab}- curl{H}_{ab}-3\sigma_{c\langle a}{E_{b\rangle}}^{c}+\frac{1}{2}\rho \sigma_{ab}&=&0,\\
\label{p:5}\dot{H}_{ab}+\Theta H_{ab}-3\sigma_{c\langle
a}H_{b\rangle}^{c}+ curl(E)_{ab}&=&0,
\end{eqnarray} and the constraints:
\begin{eqnarray}
\tilde{\nabla}^{b}{\sigma_{ab}}&=&-\frac{2}{3}\tilde{\nabla_{a}}\Theta,\\
curl(\sigma)_{ab}&=&H_{ab},\\
\tilde{\nabla}^{b}E_{ab}&=&\frac{1}{3}\tilde{\nabla}_{a}\rho+\varepsilon_{abc}{\sigma^{b}}_{d}H^{cd},\\
\tilde{\nabla}^{b}H_{ab}&=&-\varepsilon_{abc}{\sigma^{b}}_{d}E^{cd}.
\end{eqnarray}\par We linearize these equations about a dust filled Bianchi I background with zero vorticity. The background quantities are the rate of expansion ( $\Theta$), the shear tensor ($\sigma_{ab}$), the energy density ($\rho$) and the \emph{electric} part of the Weyl tensor ($E_{ab}$). Scalar-mode of perturbations are given by the conditions: 
\begin{eqnarray}
\label{pd:1}\tilde{\nabla}^{b}\rho\neq0,
\tilde{\nabla}^{b}\Theta\neq0, \tilde{\nabla}^{b}\sigma^{2}\neq0,
H_{ab}=0.\end{eqnarray} The possibility of information exchange via gravitational waves or sound is eliminated by setting $H_{ab}=0$ and $p=0$. Models with these conditions are called Silent universes \cite{bib:28}-\cite{bib:28.1}. The vanishing of the magnetic part of the Weyl tensor leads to a new constraint, $curl E_{ab}=0.$ These conditions modify the propagation and the constraints equations as follows:\begin{eqnarray}
\label{p:1}\dot{\rho}&=&-\Theta\rho,\\
\label{p:2}\dot{\Theta}&=&-\frac{1}{3}{\Theta^{2}}-\sigma_{ab}\sigma^{ab}-\small{\frac{1}{2}}\rho,\\
\label{p:3}\dot{\sigma}_{ab}&=&-\small{\frac{2}{3}}\Theta\sigma_{ab}-{\sigma_{c\langle
a}}{\sigma_{b\rangle}}^{c}-E_{ab},\\
\label{p:4}\dot{E}_{ab}&=&-\Theta E_{ab}+3\sigma_{c\langle
a}{E_{b\rangle}}^{c}-\small{\frac{1}{2}}\rho\sigma_{ab}.
\end{eqnarray}  and 
\begin{eqnarray}
\label{c:1}{C^{1}}_{a}&=&\tilde{\nabla}^{b}{\sigma_{ab}}-\frac{2}{3}\tilde{\nabla_{a}}\Theta=0,\\
\label{c:2}{C^{2}}_{ab}&=&curl \sigma_{ab}=0,\\
\label{c:3}{C^{3}}_{a}&=&\tilde{\nabla}^{b}E_{ab}-\frac{1}{3}\tilde{\nabla}_{a}\rho=0,\\
\label{c:4}{C^{4}}_{a}&=&\varepsilon_{abc}{\sigma^{b}}_{d}E^{cd}=0,\\
\label{c:5}{C^{5}}_{ab}&=& curl E_{ab}=0.
\end{eqnarray} The last constraint equation is as a result of setting $H_{ab}=0.$ We
have adopted the notation $C^{(A)}=0$ used in \cite{bib:30} where,
\begin{eqnarray}
C^{(A)}=\{\tilde{\nabla}^{b}{\sigma_{ab}}-\frac{2}{3}\tilde{\nabla_{a}}\Theta,curl\sigma_{ab}-H_{ab},..\}
\end{eqnarray} and $A=1...5$. The evolution of $C^{A}$ along
$u^{a}$ is given by the system of equations $\dot{C}^{A}=\mathcal{F}^{A}(C^{B})$ , where $\mathcal{F}^{A}$ do not have time derivatives.  Detailed constraints analysis in the $'1+3'$ formalism was given in\cite{bib:30} for the case of general nonlinear perturbations in FLRW models with dust equation of state, and in \cite{bib:31} for barotropic perfect fluid case for FLRW. In our analysis we find,
\begin{widetext}
\begin{eqnarray} \label{ce:1}\dot{C^{1}}_{a}+\Theta {C^{1}}_{a}+{C^{3}}_{a}-2{\epsilon_{a}}^{bc}{\sigma_{b}}^{d}{C^{2}}_{cd}&=&0,\\
\label{ce:2}\dot{C^{2}}_{ab}+\Theta {C^{2}}_{ab}+{\epsilon^{cd}}_{(a}{\sigma_{b)}}_{c}{C^{1}}_{d}&=&0,\\
\label{ce:3}\dot{C^{3}}_{a}+\frac{4}{3}\Theta
{C^{3}}_{a}-\frac{3}{2}{E_{a}}^{b}{C^{1}}_{b}-\frac{1}{2}{\sigma_{a}}^{b}{C^{3}}_{b}+\frac{1}{2}\rho
{C^{1}}_{a}-\frac{1}{2}curl{C^{4}}_{a}&=&\zeta_{a}^{(3)},\\
\label{ce:4}\dot{C^{4}}_{a}+\frac{5}{3}\Theta
{C^{4}}_{a}&=&\zeta_{a}^ {(4)}\\
\label{ce:5}\dot{C^{5}}_{ab}+\frac{4}{3}\Theta
{C^{5}}_{ab}-\frac{3}{2}{C^{1}}_{c(a}{E_{b)}}^{c}-\frac{3}{2}{C^{3}}_{c(a}{\sigma_{b)}}^{c}+
\frac{1}{2}\rho {C^{2}}_{ab}&=&\zeta_{ab}^{(5)}.
\end{eqnarray}
where the terms on the right hand side are new constraint conditions given by
\begin{eqnarray}
{\zeta^ {(3)}}_{a}=-{\epsilon_{a}}^{bc}({E_{b}}^{d}{C^
{2}}_{cd}-{\sigma^{d}}_{b}{C^{5}}_{cd})\\
{\zeta^{(4)}}_{a}=-\varepsilon_{abc}{\sigma_{e}}^{\langle
b}\sigma^{d\rangle e}{E^{c}}_{d}-\varepsilon_{abc}{\sigma^{b}}_{d}
{\sigma_{e}}^{\langle
c}E^{d\rangle e}\\
{\zeta^ {(5)}}_{ab}=3curl(\sigma^{c}_{\langle a}E_{b\rangle c})
-{\sigma^{e}}_{c}{\varepsilon^{cd}}_{(a}\tilde{\nabla}_{|e|}E_{b)d}-
\frac{3}{2}{\varepsilon^{cd}}_{(a}E_{b)d}\tilde{\nabla}_{e}{\sigma^{e}}_{c}-
\frac{3}{2}{\varepsilon^{cd}}_{(a}\sigma_{b)d}\tilde{\nabla}_{e}{E^{e}_{c}}.
\end{eqnarray}

Equation (\ref{ce:1}) requires (\ref{p:2}), (\ref{p:3}),
(\ref{c:1}), (\ref{c:2}), (\ref{c:3}), (\ref{app:2}), (\ref{app:6})
and (\ref{app:7}). Equation (\ref{ce:2}) requires (\ref{p:3}),
(\ref{c:1}), (\ref{c:2}), (\ref{app:3}) and (\ref{app:10}). Equation
(\ref{ce:3}) requires (\ref{p:1}), (\ref{p:3}), (\ref{p:4}),
(\ref{c:1}), (\ref{c:3}), (\ref{c:4}), (\ref{app:2}) and
(\ref{app:6}). Equation (\ref{ce:4}) requires (\ref{p:3}),
(\ref{p:4}), and (\ref{c:4}). Equation (\ref{ce:5}) requires
(\ref{app:10}), (\ref{p:4}), (\ref{c:2}), (\ref{c:4}) and
(\ref{c:5}). The point about this is that the original constraints are said to be preserved if we can demonstrate that all the new constraint vanish.

Using equation (\ref{p:3}), $\zeta_{a}^{(4)}$ can be written in the form given below. Equivalently, using equation (\ref{p:3}), the electric part of the Weyl tensor can be eliminated from $\zeta_{ab}^{(5)},$ which leads to the form given below:

\begin{eqnarray}
\label{z1}{\zeta^{(4)}}_{a}&=&
\varepsilon_{abc}[{\sigma^{b}}_{d}{\sigma_{e}}^{\langle
c}\dot{\sigma}^{d\rangle e}-{{\sigma_{e}}^{\langle
c}\sigma^{d\rangle
e}\dot{\sigma}^{b}}_{d}+{\sigma^{b}}_{d}{\sigma_{e}}^{\langle
c}{\sigma_{f}}^{\langle d\rangle}\sigma^{e \rangle
f}-{\sigma_{e}}^{\langle c}\sigma^{d\rangle e}{\sigma_{f}}^{\langle
b}{\sigma_{d\rangle}}^{f}].\\
\label{z2}{\zeta_{ab}}^{(5)}&=&\frac{3}{2}\varepsilon_{cd(a}{\sigma_{b)}}^{d}\tilde{\nabla}_{e}\dot{\sigma}^{ec}+\frac{3}{2}\varepsilon_{cd(a}{\dot{\sigma}_{b)}}^{d}\tilde{\nabla}_{e}{\sigma}^{ec}+2\Theta\varepsilon_{cd(a}{\sigma_{b)}}^{d}\tilde{\nabla}_{e}\sigma^{ec}+3\varepsilon_{cd(a}{\sigma_{b)}}^{d}\sigma^{ec}\tilde{\nabla}_{f}\sigma^{f}_{e}\nonumber\\
&&+3\varepsilon_{cd(a}{\sigma_{b)}}^{d}\tilde{\nabla}_{e}(\sigma^{f\langle
c}{\sigma^{e\rangle}}_{f})+{\sigma_{e}}^{c}\varepsilon_{cd\langle
a}\tilde{\nabla}^{e}{\dot{\sigma}_{b\rangle}}^{d}+\frac{2}{3}\Theta{\sigma_{e}}^{c}\varepsilon_{cd\langle
a}\tilde{\nabla}^{e}{{\sigma}_{b\rangle}}^{d}+{\sigma_{e}}^{c}\varepsilon_{cd\langle
a}{\sigma_{b\rangle}}^{d}\tilde{\nabla}^{f}{\sigma^{e}}_{f}\nonumber\\
&&+{\sigma_{e}}^{c}\varepsilon_{cd\langle
a}\tilde{\nabla}^{e}({\sigma_{b\rangle
f}\sigma^{df})+\frac{3}{2}\varepsilon_{cd\langle
a}{\sigma_{b\rangle}}^{c}\tilde{\nabla}_{e}\dot{\sigma}^{ec}+\frac{3}{2}\varepsilon_{cd\langle
a}{\dot{\sigma}_{e\rangle}}^{c}\tilde{\nabla}^{e}{\sigma_{b\rangle}}^{d}+\Theta\varepsilon_{cd(a}{\sigma_{b)}}^{d}\tilde{\nabla}_{e}{\sigma_{b\rangle}}^{c}}+\Theta\varepsilon_{cd(a}{\sigma_{b)}}^{d}\tilde{\nabla}_{e}\sigma^{ec}\nonumber\\
&&+\Theta\varepsilon_{cd\langle
a}\sigma^{ec}\tilde{\nabla}_{e}{\sigma_{b\rangle}}^{d}+\frac{3}{2}\varepsilon_{cd(a}{\sigma_{b)}}^{d}\tilde{\nabla}_{e}\sigma^{fe}\sigma^{df}+\frac{3}{2}\varepsilon_{cd(a}\sigma^{fc}\sigma^{fe}\tilde{\nabla}_{e}{\sigma_{b\rangle}}^{d}.
\end{eqnarray} 
\end{widetext} It can be shown that the terms in equation \ref{z1} cancel each other out.  The fully expanded form of $\zeta_{ab}^{(5)}$ given by equation \ref{z2}
${\zeta^{(5)}}_{ab}$ is equivalent to equation (15) in \cite{bib:31}, where it is shown there that the term vanishes when the shear is diagonal and degenerate. We note that our equation \ref{c:4} shows that one can choose a common eigen-frame for both shear and the electric part of the Weyl tensor, and one that leads to a simultaneous diagonalization of both. Such an eigen-frame is Fermi transported along the 4-velocity. This can be achieved by setting the off-diagonal terms of the shear and the electric part of the Weyl tensor set to zero. It is then possible to choose a tracefree-adapted irreducible frame components for $\sigma_{ab}$ and $E_{ab}$ defined by \[\sigma_{+}:=-\frac{1}{2}\sigma_{11}=\frac{1}{2}(\sigma_{22}+\sigma_{33}), \sigma_{-}:=\frac{1}{2\surd{3}}(\sigma_{22}-\sigma_{33}),\] and where vorticity is zero. $E_{+/-}$ is defined in a similar manner. $\zeta_{ab}^{(5)}$ now takes a form similar to equations 67-69 in \cite{bib:31}. From our equation (\ref{c:4}), ($\sigma_{-}=0\Rightarrow E_{-}=0$). It is then straightforward to show, using the tetrad formalism, that $\sigma_{-}=0(i.e.\sigma_{22}=\sigma_{33})$ leads to $\zeta_{ab}^{(5)}=0$. This means that, subject to the preceding discussions, all the new constraints vanish leading the preservation of the original constraints.
This implies that the conditions given above for pure, or decoupled, density perturbations about a dust filled Bianchi I model with zero vorticity represent consistent characterization only when the background shear tensor is degenerate.

It is important to point out that this is not a proof that the solutions to the constraints exist Ñ only that if solutions exist, then they evolve consistently. The question of existence is one that needs to be investigated. Such an investigation should show how to construct a metric from given initial data in the covariant formalism. This notwithstanding, one can now study the decoupled perturbations as will be presented in \cite{bib:36}.

The following commutation relations for a scalar quantity, {\it f} and PSTF-tensor quantities, \(T_{ab}\) and \(V_{ab}\)  with vanishing vorticity have been used in our analysis;
\begin{widetext}
\begin{eqnarray}
\label{app:0}\varepsilon_{abc}\varepsilon^{dec}&=&2!{h^{d}}_{[a}{h^
{e}}_{b]},\\
\label{app:1}\varepsilon_{abc}{T^{b}}_{p}{T^{p}}_{q}V^{cq}&=&-T_{ab}
\varepsilon^{bcd}{T_{c}}^{p}V_{dp},
\\
\label{app:2}(\tilde{\nabla}_{a}f)^{.}&=&\tilde{\nabla}_{a}\dot{f}-
\frac{1}{3}\Theta\tilde{\nabla}_{a}f-{\sigma_{a}}^{b}\tilde{\nabla}_{b}f,
\\
\label{app:4}\tilde{\nabla}_{[a}\tilde{\nabla}_{b]}f&=&0,
\\
\label{app:3}curl(T^{2})_{ab}&=&\varepsilon_{cd(a}\tilde{\nabla}^{e}\{{T_{b)}^{c}{T^{d}}_{e}\}},
\\
\label{app:7}\varepsilon_{abc}{T^{b}}_{d} curl
V^{cd}&=&T^{bc}\tilde{\nabla}_{a}V_{bc}-T^{bc}\tilde{\nabla}_{b}V_
{ac}-\frac{1}{2}T_{ab}\tilde{\nabla}_{c}V^{bc},\nonumber\\
\\
\label{app:8}curl(f T_{ab})&=&f
curl(T)_{ab}+\varepsilon_{cd(a}T_{b)}^
{d}\tilde{\nabla}^{c}f,\\
\label{app:5}(\tilde{\nabla}_{a}T_{bc})^{.}&=&\tilde{\nabla}_{a}\dot{T}_{bc}-\frac{1}{3}\Theta\tilde{\nabla}_{a}T_{bc}-{\sigma_{a}}^{d}\tilde{\nabla}_{d}T_{bc}+2{H_{a}}^{d}\varepsilon_{de(b}T_{c)}^{e}
,\\
\label{app:6}(\tilde{\nabla}^{b}T_{ab})^{.}&=&\tilde{\nabla}^{b}\dot
{T}_{ab}-\frac{1}{3}\Theta\tilde{\nabla}^{b}T_{ab}-\sigma^{bc}\tilde{\nabla}_
{c}T_{ab}+\varepsilon_{abc}{H^{b}}_{d}T^{bc},
\\
\label{app:9}(\tilde{\nabla}^{b}curl
T_{ab})&=&\frac{1}{2}\varepsilon_
{abc}\tilde{\nabla}^{b}\tilde{\nabla}_{d}T^{cd}+\varepsilon_{abc}{T^
{b}}_{d}(\frac{1}{3}\Theta\sigma^{cd}-E^{cd})-\sigma_{ab}\varepsilon^{bcd}\sigma_{ce}{T^{e}}{d},\\
\label{app:10}(curl T_{ab})^{.}&=&
curl(\dot{T})_{ab}-\frac{1}{3}\Theta curl
T_{ab}-{\sigma_{e}}^{c}\varepsilon_{cd(a}\tilde{\nabla}^{e}{T_{b)}}^
{d}+3H_{c\langle a}{T_{b\rangle}}^{c},
\\
\label{app:11}curl
curl(T)_{ab}&=&-\tilde{\nabla}^{2}{T}_{ab}+\frac{3}
{2}\tilde{\nabla}_{\langle a}\tilde{\nabla}^{c}T_{b\rangle
c}+(\rho-\frac{1}{3}\Theta^{2})T_{ab}+\nonumber\\&&3T_{c\langle
a}\{{E_{b\rangle}}^{c}-\frac{1}{3}\Theta{\sigma_{b\rangle}}^{c}\}+\sigma_{cd}T^{cd}\sigma_{ab}-T^{cd}\sigma_{ca}\sigma_{bd}+\sigma^{cd}\sigma_{c(a}T_{b)d}.
\end{eqnarray}
\end{widetext}
This work was funded URC grant from the University of Cape Town.


\begin{thebibliography}{40}
\bibitem{bib:1} Oliveria-Costa A et al 2004 Phys. Rev. D 69 063515
\bibitem{bib:2} Schwarz D et al 2004 Phys. Rev. Lett. 93 221301
\bibitem{bib:3} Ralston J and Jain, 2004 Int. J. Mod. Phys. 13 1857
\bibitem{bib:4} Eriksen H K et al 2004 Astrophys. J. 14  605
\bibitem{bib:5} Eriksen H K et al 2004 Astrophys. J. 64 
\bibitem{bib:51}Eriksen H K et al 2005 Astrophys. J.622 58
\bibitem{bib:7} Hansen F K et al 2004 Mon. Not. Roy. Astron. Soc. 354: 641
\bibitem{bib:8} Land K and Magueijo J 2005 Mon. Not. Roy. Astron. Soc. 357 994
\bibitem{bib:9} Hansen F K et al, 2004 Astrophys. J. 607 L67
\bibitem{bib:10} Vielva P et al 2004 Astrophys. J. 609 22
\bibitem{bib:Planck} Ade P A R 2013 Astro. \& Astrophys.
\bibitem{bib:10.1} Bennett C L et al 2003 Astrophys. J. Suppl. 148 1
\bibitem{bib:11} Land K and Magueijo J 2005 Phys. Rev. Lett. 95 071301
\bibitem{bib:12} Lifshitz E M 1946 J. Phys (USSR). 10 116
\bibitem{bib:13} Bonnor W B 1957  Mon. Not. Roy. Astron. Soc. 117 104
\bibitem{bib:20} Ellis G F R  and Bruni M 1990 Phys. Rev. D 40 1804
\bibitem{bib:21} Ellis G F R, Bruni M and Hwang J 1990 Phys. Rev. D 42 1035
\bibitem{bib:222} Ellis G F R 1971 General Relativity and Cosmology (New York: Academic)
\bibitem{bib:23} Perko T E, Matzner R A and Shepley L C 1972 Phys. Rev. D 6 969
\bibitem{bib:24} Tomita K and Mitsue D 1986 Phys. Rev. D 34 3570
\bibitem{bib:26} Heyrim N and Hwang J 1995 Phys. Rev. D 52 5643-52 : Phys. Rev. D 52 1970: Phys. Rev. D 53 4311
\bibitem{bib:26.1} Ellis G F R  and van Elst H 1999, NATO Adv. Study Inst. Ser. C. Math. Phys. Sci. 541
\bibitem{bib:28} Matarrese S, Pantano O and Saez D 1994 Phys. Rev. Lett. 72 320
\bibitem{bib:28.1}Bruni M, Matarrese S and Pantano O 1995 Astrophys. J. 445 958
\bibitem{bib:30} Maartens R1997 Phys. Rev.D 55
\bibitem{bib:31} van Elst H 2013 {http://www.maths.qmul.ac.uk/~hve/13coveq.pdf}.
\bibitem{bib:35} Dunsby P K S 1993 Phys. Rev. D 48  3562
\bibitem{bib:36} Osano B 2013 (in preparation)
\end{thebibliography}
\end{document}